\documentclass[journal, final]{IEEEtran}

\usepackage{kotex}
\usepackage{amsmath,graphicx}
\usepackage{amsfonts}
\usepackage[ruled]{algorithm2e}
\usepackage{algorithmic}
\usepackage{cite,url}
\usepackage{multirow}
\usepackage{array}
\newcolumntype{P}[1]{>{\centering\arraybackslash}p{#1}} \usepackage{tipa} 
\usepackage{booktabs}
\usepackage[flushleft]{threeparttable}
\usepackage{float}
\usepackage[caption=false]{subfig}
\usepackage{epstopdf}

\newcommand{\colvec}[2][.8]{
 \scalebox{#1}{
  \renewcommand{\arraystretch}{.8}
  $\begin{bmatrix}#2\end{bmatrix}$
 }
}
\newcommand{\colvecc}[2][.8]{
 \scalebox{#1}{
  \renewcommand{\arraystretch}{.8}
  $\begin{pmatrix}#2\end{pmatrix}$
 }
}
\newcommand{\colveccc}[2][.8]{
 \scalebox{#1}{
  \renewcommand{\arraystretch}{.8}
  $#2$
 }
}

\newcommand\ttt{\rule{0pt}{2.6ex}}    
\newcommand\bbb{\rule[-1.2ex]{0pt}{0pt}}

\begin{document}

\title{Lyrics-to-Audio~Alignment by Unsupervised~Discovery of Repetitive~Patterns in~Vowel~Acoustics}

\author{Sungkyun~Chang,~\IEEEmembership{Member,~IEEE,}
        and~Kyogu~Lee,~\IEEEmembership{Senior Member,~IEEE}
\thanks{Manuscript created 18, January 2017. This work was supported by a National Research Foundation of Korea (NRF) grant funded by the MSIP (NRF-2014R1A2A2A04002619).}%
\thanks{The authors are with Music and Audio Research Group (MARG), Department of Transdisciplinary Studies, Seoul National University, Seoul 08826, Korea, e-mail: \{rayno1, kglee\}@snu.ac.kr}%
\thanks{Kyogu Lee is also with the Advanced Institutes of Convergence Technology, Suwon, Korea.}}%

\markboth{First Draft, 18 January 2017}%
{Chang \MakeLowercase{\textit{et al.}}: Lyrics-to-Audio Alignment by Unsupervised Discovery of Repetitive Patterns in Vowel Acoustics}
\maketitle

\begin{abstract}
Most of the previous approaches to lyrics-to-audio alignment used a pre-developed automatic speech recognition (ASR) system that innately suffered from several difficulties to adapt the speech model to individual singers. A significant aspect missing in previous works is the self-learnability of repetitive vowel patterns in the singing voice, where the vowel part used is more consistent than the consonant part. Based on this, our system first learns a discriminative subspace of vowel sequences, based on weighted symmetric non-negative matrix factorization (WS-NMF), by taking the self-similarity of a standard acoustic feature as an input. Then, we make use of canonical time warping (CTW), derived from a recent computer vision technique, to find an optimal spatiotemporal transformation between the text and the acoustic sequences. Experiments with Korean and English data sets showed that deploying this method after a pre-developed, unsupervised, singing source separation achieved more promising results than other state-of-the-art unsupervised approaches and an existing ASR-based system.
\end{abstract}



\section{Introduction}
\label{sec:Intro}
\IEEEPARstart{I}{n} the last decade, there has been considerable interest in digital music services that display the lyrics of songs that are synchronized with their audio. As a recent example, {\it Sound Hound}\footnote{https://www.soundhound.com/} provided a live-lyrics feature using an audio fingerprint\cite{Wang:2006:SMR:1145287.1145312} technique, with known time stamps for alignment. Although some well-known hit songs already have lyric time stamps within existing karaoke databases, more than a hundred million numbers, including new, unpopular songs {\it YouTube}\footnote{https://www.youtube.com/}, cover songs, and live recordings may not. An automatic {\it lyrics-to-audio} alignment system could reduce the huge amount of time and labor required to manually construct such time stamps. Not limited to the application above, it could also be used as a front end for various purposes in content-based music information retrieval (MIR) involving lyrical rhymes \cite{mayer2008rhyme} and emotions \cite{mcvicar2011mining, hu2011exploring}.
Previous works on automatic {\it lyrics-to-audio} alignment systems for popular music recordings can be classified into three categories.

\subsubsection{ASR-based}
The core part of these systems \cite{fujihara2006automatic, fujihara2011lyricsynchronizer, mauch2012integrating, McVicar2014} uses mature automatic speech recognition (ASR) techniques developed previously for regular speech input. The phone model is constructed as gender-dependent models through training with a large speech corpus \cite{kawahara2004recent}. In parallel, the music recordings are pre-processed by a singing voice separation algorithm to minimize the effect of unnecessary instrumental accompaniment. In this situation, adapting the trained speech model to the segregated sung voice is a difficult problem, due to the large variation in individual singing styles, remaining accompaniments, and artifact noises. Based on the standard ASR technique, Fujihara, et al. \cite{fujihara2006automatic,fujihara2011lyricsynchronizer} proposed a three-step adaptation strategy: i) clean speech to clean singing, ii) clean singing to segregated singing, and iii) segregate speech to a specific singer. Then a {\it Viterbi} alignment, based on left-to-right HMM\cite{18626} architecture, searches for the optimal path of lyrics-to-audio synchronization. Fujihara's initial\cite{fujihara2006automatic} and final works\cite{fujihara2011lyricsynchronizer} were evaluated in terms of sentence-level alignments of Japanese popular music. Later, their final work\cite{fujihara2011lyricsynchronizer}---referred to as the baseline in the present research---achieved a word-level accuracy of 46.4\% \cite{mauch2012integrating} in English popular music, defining error as the number of word displacements over 1 s.

\subsubsection{Musical knowledge-based}
The use of music structural information from chord analysis has provided good guidance for higher-level alignments in previous works\cite{lee2008segmentation, Kan2008Lyrically}. Lee et al. \cite{lee2008segmentation} showed that the results of rough structure segmentation with hand-labeled paragraphs in lyrics can be used for paragraph-level alignment. To deal with sentence-level alignment, Kan, et al.\cite{Kan2008Lyrically} proposed a system named {\it Lyrically}: By assuming that only verse and chorus parts would contain sung voice, it first performed a chorus-detection using structure segmentation to find voice regions. Then, they used a heuristic algorithm, based on a rough estimation of each sentence's text duration, to allocate lines of text to audio segments. Thus, a major limitation of {\it Lyrically} is that it can only align verse and chorus sections, while vocals in the bridge and outro sections are not considered. 

\subsubsection{Combined approach}
With respect to word-level alignment, Mauch et al.\cite{mauch2012integrating} proposed the integration of a chord-to-audio alignment method into Fujihara's ASR-based system \cite{fujihara2011lyricsynchronizer}. This greatly improved the word-level accuracy of the ASR-based system\cite{fujihara2011lyricsynchronizer}, from 46.4\% to 87.5\%. However, the requirements of Mauch's approach are strongly limiting when applied to real-world problems; for each unseen song, the suggested method requires a manually transcribed score that contains chord labels with reliable temporal information corresponding to the lyrics.

\begin{figure}
\centering
\subfloat[Overview of our system. \label{fig:system_overview_sub1}]{%
       \includegraphics[width=1\columnwidth]{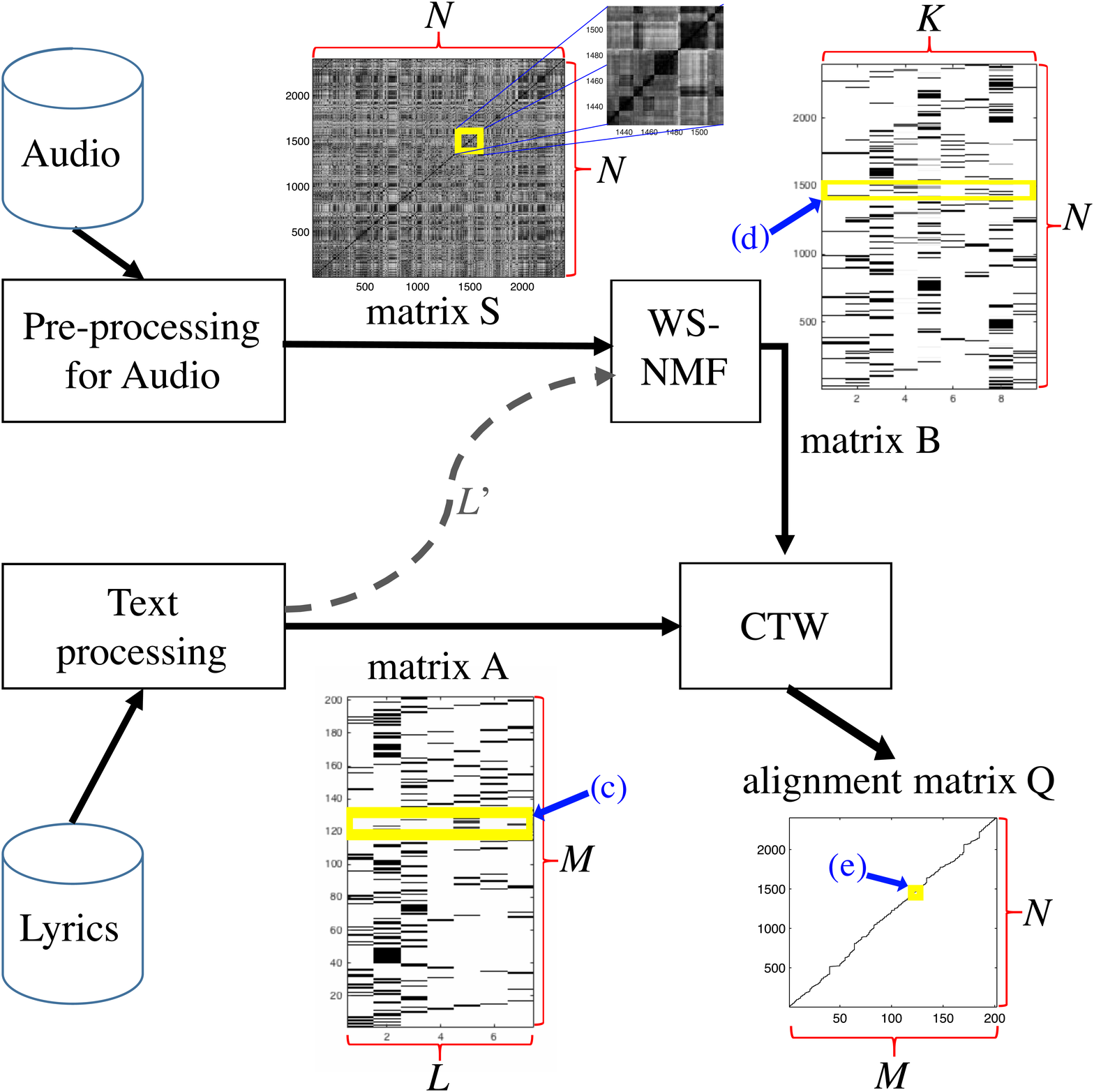}
     }
\\
\subfloat[A vowel sequence, corresponding to the yellow box area of (a).  \label{fig:system_overview_sub2}]{%
       \includegraphics[width=.9\columnwidth]{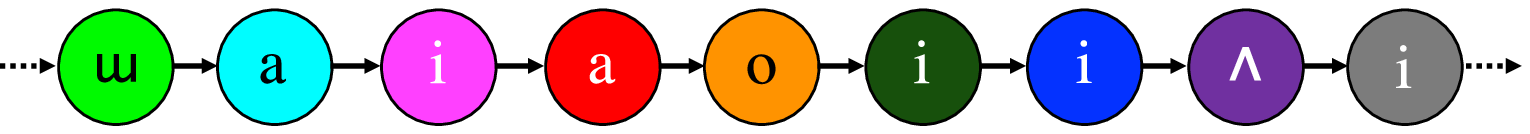}
     }
\\
\subfloat[Magnification of matrix $A$. \label{fig:system_overview_sub4}]{%
       \includegraphics[width=0.275\columnwidth]{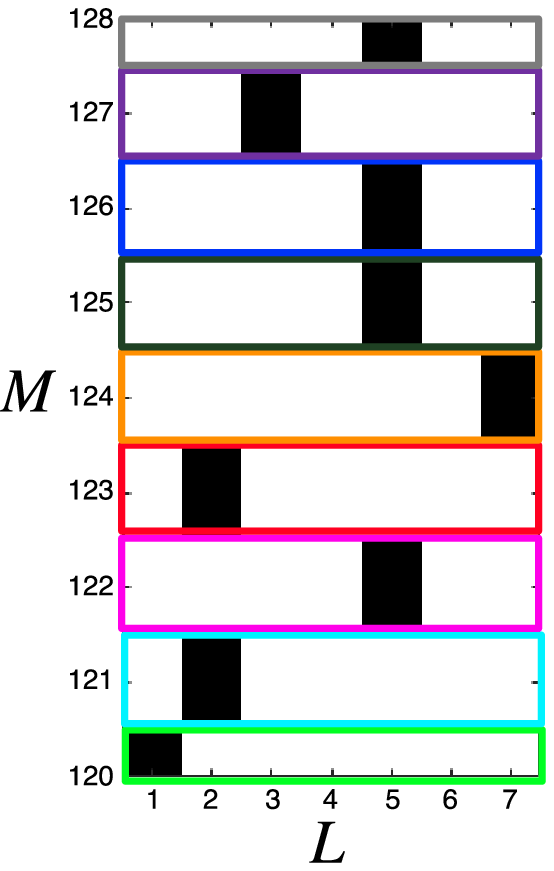}
     } \hspace{0.0056\columnwidth} 
\subfloat[Magnification of matrix $B$. \label{fig:system_overview_sub5}]{%
       \includegraphics[width=0.317\columnwidth]{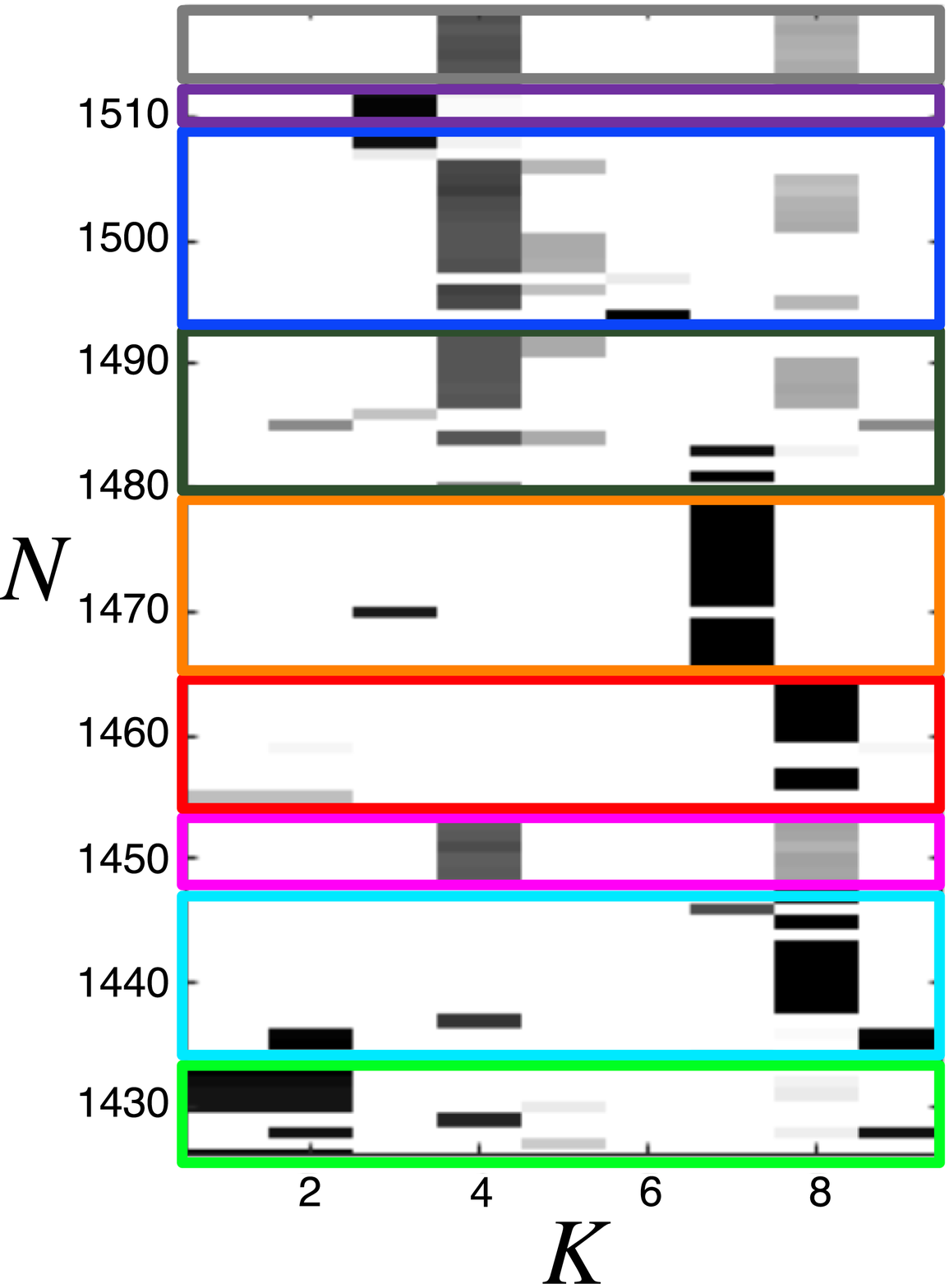}
     } \hspace{0.0056\columnwidth}
\subfloat[Magnification of alignment mapping matrix $Q$. \label{fig:system_overview_sub6}]{%
       \includegraphics[width=0.330\columnwidth]{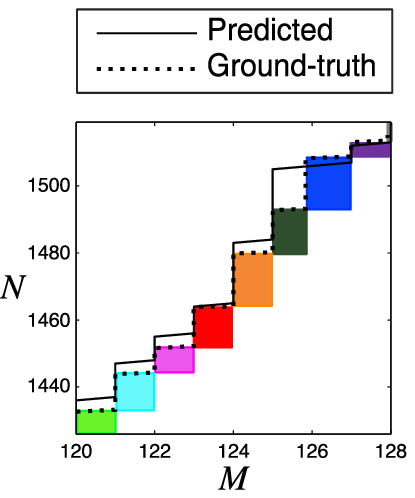}
     }  
     
\caption[caption]{In (a), the matrix $S$ represents the self-similarity within $N$ samples observed in the audio, while the matrix $A$ from the textual lyrics is a logical time series matrix that represents assignments between the $L$ classes of vowels and the $M$ lengths of time series. The proposed WS-NMF+CTW consists of two sequentially organized methods: WS-NMF first takes matrix $S$, and produces matrix $B$, which represents unlabeled vowel activations with the size of $N \times K$, where $K \times K$ with $K\approx L'$ represents the size of a latent vowel subspace $W$ in each song. Then, CTW performs an alignment between matrixes $A$ and $B$. The resulting matrix $Q$ from CTW represents a predicted assignment between the $N$th samples in the audio and the $M$th syllable of the lyrics. Sub-figures from (c) to (f) demonstrate the data transformations. They focus on the yellow box area in (a), which corresponds to the vowel sequence displayed in (b). In (d), (e), and (f), we displayed a ground-truth mapping between the vowel and audio sequences by using colors defined in (b). For visualization of matrixes, we used a song, ``Ahn Ye Eun -  Sticker''.} 

\label{fig:system_overview}
\end{figure}

A previous work not listed above is a supervised learning approach, based on a source-filter model and non-negative {\it matrix factorization}\cite{lee1999learning}. Pedone et al.\cite{pedone2011phoneme} makes the assumption of representing the sung voice as a convolution of generative glottal sources and a trainable filter. The system achieved an F-measure of 43\% for syllable-level alignment. However, their experimental conditions are regarded as a limited case, in that it was tested with vocal-boosted synthesized audio tracks.

In this study, we sought a more generic and efficient method to deal with both syllable- and word-level {\it lyrics-to-audio} alignments of music recordings. Our approach was to discover repetitive acoustic patterns of vowels in the target audio by referencing vowel patterns appearing in lyrics. Thus, the present method does not require any external training data for constructing a specific model for speech or its adaptation to specific singers, unlike the previous approaches\cite{fujihara2011lyricsynchronizer, mauch2012integrating}. Moreover, the present work does not depend on a manually transcribed chord score that was required in some previous work\cite{mauch2012integrating}. Instead, we make full use only of information derived from the target audio and lyrics. In doing so, we can expect the advantage of avoiding the problems above that occur due to the large variation in singing styles and singers' voices in different songs. Furthermore, we show that the present method is applicable to two languages, with many fewer modifications in experiments.

The remaining sections are outlined as follows. An overview of the proposed system is provided in Figure \ref{fig:system_overview}. Section \ref{sec:preliminaries} first describes our key idea from the nature of the sung voice. Then, we explain several necessary components for audio pre-processing. Text processing is explained in Section \ref{sec:textProcessing}. Section \ref{sec:vowelSubspaceLearn} proposes a discriminative subspace learning method, assisted by a variant of the {\it matrix factorization} technique \cite{ding2005equivalence} to capture repetitive vowel patterns from a self-similarity matrix. Section \ref{sec:spatialtemporal_ctw} describes a method to align matrixes belonging to two different domains, in terms of a spatiotemporal transformation. Section \ref{sec:evaluation}--\ref{sec:result} compares and analyzes the performance of the present work with other methods, using two data sets. Then, we provide conclusions in Section \ref{sec:print}.

\section{Preliminaries}
\label{sec:preliminaries}
\begin{figure}[t]
	\centerline{{
	\includegraphics[width=1\columnwidth]{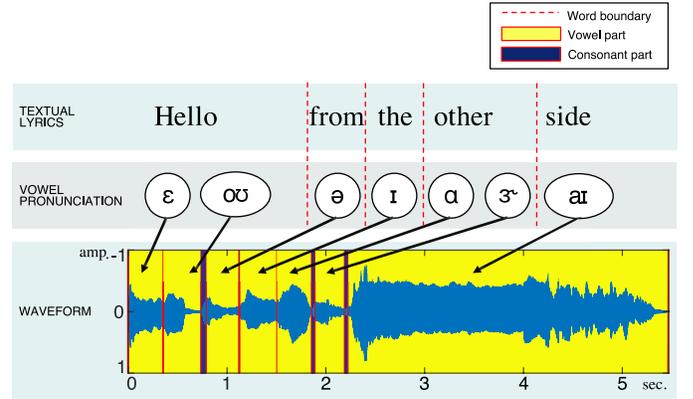}}}
	\caption{Examples of elongated vowels in the singing voice of a popular song, ``Adele - Hello''. The top line displays a snippet of the textual lyrics. The middle line displays the vowel pronunciation symbols (in IPA) of words retrieved from a dictionary. The bottom line shows the distribution of vowels. Each consonant and vowel part is colored in dark blue and yellow, respectively.}
 	\label{fig:example1}
\end{figure}

\subsection{Assumption---`Observing Only Vowels'}
\label{ssec:Assumption}
Singing can be considered as a type of speaking, although it deviates significantly from regular speech. A common property of a traditional singing voice is that the duration of a sung vowel is typically much longer than that of other syllabic components: according to a previous study \cite{mabry2002exploring}, such elongation and purity of vowel formation in singing has been required for singers to sustain a tone and pitch consistently. This feature intuitively leads us to an assumption that {\it lyrics-to-audio} alignment may be dealt with by observing only dominant vowel parts within a singing voice. Figure \ref{fig:example1} gives an example of how vowel and consonant parts are distributed in audio sample waveforms of popular music recordings. A similar assumption---alignment by observing only vowels---was made initially by Fujihara\cite{fujihara2006automatic}. Because overlooking this assumption may yield a severe bottleneck problem in the generalization of our model, we need to validate the assumption. This issue is covered in Section \ref{ssec:Validation_vowel_only}, in terms of the theoretical upper bound of alignment performance.

\subsection{Pre-processing of Audio}
\label{subsec:Pre-processing_audio}
Because our system discovers repetitive vowel patterns by observing audio similarity, an important prerequisite is to reduce unwanted effects of accompaniment parts that may disturb the observation of sung vowel audio signals. For this reason, we pre-process input audio signals with singing voice separation and voice activity detection (VAD) algorithms. From the segregated singing source signals in voice-activated regions, standard audio features are extracted. A self-similarity matrix (SSM), $S$, is then generated from the samples of features. In summary, we chain multiple pre-processors: 1) singing voice separation$\rightarrow$ 2) VAD $\rightarrow$ 3) feature extraction$\rightarrow$ 4) generation of a SSM.

\subsubsection{Singing Voice Separation}
We used a low-rank source separation method based on {\it robust principal component analysis} (RPCA), as proposed by Huang\cite{svad_Huang2012}. Here, the accompaniment parts can be assumed to be in a low-rank subspace in RPCA, while voice parts are assumed to be relatively sparse. Using a spectrogram as an input, the RPCA-based method generates a binary time-frequency mask for separation of the singing voice. We then reconstruct the singing voice signal from the processed spectrogram. The parameters for RPCA were set at \{16,000 Hz, 1,024 samples, 256 samples, and 1\} for the audio sampling rate, window size, hop size, and the regularization parameter $\lambda$, respectively.

\subsubsection{VAD}
\label{ssec:VAD_RPCA}
Although some recent supervised VAD algorithms\cite{lehner2013towards, leglaive2015singing} have reported F-measures for singing voice activity detection exceeding 85\%, we found that one such algorithm\cite{lehner2013towards} tested (see Section \ref{ssec:effect_of_VAD}) with our data set was not sufficiently robust. As an alternative, we simply used the log-energy of the singing voice signal segregated by RPCA as a detection function for voice activity. Here, voice-active or -inactive frames with a 32-ms window were classified by a hard thresholding function with a global parameter $\theta$. Then, we applied smoothing (7-order median filtering) to the decision output by selecting the majority of the total votes for voice activity within 224 ms around the center frame.
The suggested parameter $\theta=$1.88 was the value chosen by a nested grid-search for parameter selection with the present VAD, which will be described along with the benchmark test in Section \ref{ssec:effect_of_VAD}.

\subsubsection{Feature Extraction}
From the accompaniment-reduced and voice-detected audio frames, we then extracted {\it mel-frequency cepstral coefficients} (MFCCs) that can compactly represent the special envelopes of vowel features. For the extraction of MFCCs, we use a set of parameters, $\{16,000 \text{~Hz}, 1024 \text{~samples~}(64 \text{~ms}), 1024 \text{~samples}, \text{~and~} 13\}$ for the audio sampling rate, window size, hop size, and the number of dimensions, respectively. Then, we took the remnant 12 dimensional feature vectors by reducing the first dimension of the MFCCs.

\subsubsection{Generation of SSM}
To measure the inter-frame acoustic similarity from the feature vectors, we generated a self-similarity matrix S, defined as 

\begin{equation}
S_{i,j}=\frac{\text{max}(\Delta)-\Delta_{i,j}}{\text{max}(\Delta)},\hspace*{.2cm} \text{s.t.}\hspace*{.15cm} \Delta_{\{i,j|i,j\in N\}}= \delta(y_i,y_j), 
\label{eq:ssm_gen}
\end{equation}
where $N, y_i$, and $\delta$ are the frame index, the $i$th frame of the MFCCs vector, and a distance function (also known as the {\it heat kernel}), respectively. Here, the distance function $\delta$ is defined as

\begin{equation}
	\delta(y_i,y_j) = e^{-z}, \text{ s.t. } z = \frac{{\parallel y_i-y_j \parallel}^2}{\sigma} .
	\label{eq:heat_kernel} 
\end{equation}

Note that in Equation \ref{eq:ssm_gen}, we typically divide the upper term by the maximum distance $\text{max}(\Delta)$, to normalize the matrix $S$ to have a non-negative value range, between 0 and 1.

\section{Text Processing}
\label{sec:textProcessing}
The goal of the text processing is to generate a vowel sequence matrix from the textual lyrics of each target song. Assume a sequence allowing repetition and that we must pick one vowel each time. The sequence $P$ with time index $m$, where $m=\{1,2,\ldots,M\}$, can be denoted as

\begin{equation}
	P_{m} = \{p_1,p_2,\ldots,p_M\},
\end{equation}
where $M$ is the total length of the sequence. Based on Table \ref{table:vowel_classes}, we define $L$ classes ($L_{\text{kr}}$ for Korean, or $L_{\text{eg}}$ for English) of vowels with label $l$ as \begin{math} l = \{1,2,\ldots,L\} \end{math}.

Then, the text processor can generate a logical matrix $A$ with $A \in \mathbb{R}^{M \times L}$,

\begin{equation}
	A_{m,l} = 
\begin{cases}
   1 & \text{if } p_{m} = l\\
   0 & \text{otherwise,}
\end{cases}
\end{equation}
as an output. Thus, the sum of elements in each row must be 1. The following illustrates details of extracting vowels from textual lyrics in two languages.

\begin{table}[t]
\caption{Sampling of IPA\cite{international1999handbook} symbolized Vowel Classes:\hspace{\textwidth} $L_{\text{kr}}=7$ for Korean, $L_{\text{eg}}=15$ for English.} 
\centering
		{
		\begin{tabular}{P{1.5cm} P{2cm} P{1cm} P{2cm} } 

			\toprule
			Language & Group & Vowel Class & Example \\
			\midrule
			\multirow{7}{*}{Korean ({\it hangul})} & \multirow{7}{*}{-} & \multicolumn{1}{c}{\textipa{a}} & \multicolumn{1}{c}{ㅏ}\\ 
                 && \multicolumn{1}{c}{\textipa{e}} & \multicolumn{1}{c}{ㅔ}\\
                 && \multicolumn{1}{c}{\textipa{i}} & \multicolumn{1}{c}{ㅣ}\\
                 && \multicolumn{1}{c}{\textipa{o}} & \multicolumn{1}{c}{ㅗ}\\
                 && \multicolumn{1}{c}{\textipa{u}} & \multicolumn{1}{c}{ㅜ}\\
                 && \multicolumn{1}{c}{\textipa{2}} & \multicolumn{1}{c}{ㅓ}\\                 && \multicolumn{1}{c}{\textipa{W}} & \multicolumn{1}{c}{ㅡ} \bbb \\
      \midrule
			\multirow{15}{*}{English} & \multirow{10}{*}{monophthongs} & \multicolumn{1}{c}{\textipa{O}} & \multicolumn{1}{c}{off } \\ 
                 && \multicolumn{1}{c}{\textipa{A}} & \multicolumn{1}{c}{far}\\
                 && \multicolumn{1}{c}{i} & \multicolumn{1}{c}{she}\\
                 && \multicolumn{1}{c}{u} & \multicolumn{1}{c}{you}\\
                 && \multicolumn{1}{c}{\textipa{E}} & \multicolumn{1}{c}{red}\\
                 && \multicolumn{1}{c}{\textipa{I}} & \multicolumn{1}{c}{pig}\\
                 && \multicolumn{1}{c}{\textipa{U}} & \multicolumn{1}{c}{should}\\
                 && \multicolumn{1}{c}{\textipa{2}} & \multicolumn{1}{c}{but}\\
                 && \multicolumn{1}{c}{\textipa{9}} & \multicolumn{1}{c}{sofa}\\
                 && \multicolumn{1}{c}{\textipa{ae}} & \multicolumn{1}{c}{at} \bbb \\ \cline{2-4}
                 &\multirow{5}{*}{diphthongs} & \multicolumn{1}{c}{e\textipa{I}} & \multicolumn{1}{c}{day} \ttt \\
                 && \multicolumn{1}{c}{a\textipa{I}} & \multicolumn{1}{c}{my}\\
                 && \multicolumn{1}{c}{o\textipa{U}} & \multicolumn{1}{c}{low}\\
                 && \multicolumn{1}{c}{a\textipa{U}} & \multicolumn{1}{c}{now}\\                 && \multicolumn{1}{c}{\textipa{O}\textipa{I}} & \multicolumn{1}{c}{boy} \bbb \\
			\bottomrule
		\end{tabular}
		}
\label{table:vowel_classes}
\end{table}

\subsection{Korean ({\it hangul})}
The generation of a vowel sequence matrix from text written in Korean letters---also known as {\it hangul}---is straightforward in that vowel pronunciation never varies within different words. Generally, each {\it hangul} letter has the form of a grouped syllable, consisting of initial, medial, and (optionally) final parts. From UTF-8-formatted text data, we can extract vowel parts by taking only medial parts. 
We define $L_{\text{kr}}$($=7$) classes of Korean vowels represented by the {\it international phonetic alphabet} (IPA) \cite{international1999handbook}, as shown in Table \ref{table:vowel_classes}. Due to the {\it phonogramic} feature of {\it hangul}, we can convert the extracted vowel parts of {\it hangul} text directly into a logical sequence matrix $A$ without the need for an external pronunciation dictionary.

\subsection{English}
In English, a certain vowel alphabet can often have a different sound within words. For example, the sounds of the letter ``a'' in words such as ``apple'' and ``car'' are very different.
To retrieve practical vowel phonetics of English words, we used a machine-readable dictionary, developed previously in CMU\cite{weide_cmudict}. The text processor first splits each word from text strings by spacing. After searching for word pronunciation from the dictionary, it reduces the consonant parts simply by referencing a vowel class table. The retrieved vowel pronunciation symbols fall into one of the $L_{\text{eg}}$($=15$) classes of vowels, which include 10 {\it monophthongs} and 5 {\it diphthongs}, as shown in Table \ref{table:vowel_classes}. By referencing the table, it is possible to convert the extracted vowel parts directly into a logical sequence matrix $A$.

\section{Unsupervised Vowel Subspace Learning with WS-NMF}
\label{sec:vowelSubspaceLearn}
This section describes a method, derived from non-negative {\it matrix factorization} (NMF), that can capture repetitive patterns and search discriminative subspace in SSM, denoted as $S$, produced by the preprocessors described in the previous section.

\subsection{Weighted, Symmetric NMF}
\label{ssec:cnmf}
The well-known standard NMF \cite{lee1999learning} can be defined as 

\begin{equation}
X_{+}\approx F_{+}G_{+}^{T},
\label{eq:stdnmf}
\end{equation} 
where $F$ and $G$ are non-negative (denoted by subscript $+$) matrixes that factorize a given non-negative matrix $X$.
Its various extensions have been applied to numerous problems in DNA analysis, image processing, and audio source separation, as examples\cite{cichocki2009nonnegative}.
Then, the symmetric NMF \cite{ding2005equivalence, cichocki2009nonnegative} can be defined as 

\begin{equation}
X_{+}\approx G_{+}G_{+}^{T}.
\label{eq:symnmf}
\end{equation} 

Thus, we regard this as a special type of NMF, where $F=G$. Over the past decade, variants of symmetric NMF have been used as scalable data analysis tools\cite{yang2012clustering}.

The weighted, symmetric (WS-) NMF \cite{ding2005equivalence} is defined as

\begin{equation}
 X_{+}\approx B_{+}WB_{+}^{T},
 \label{eq:wenmf}
\end{equation} 
where the non-negative matrix $X$ used to be a {\it positive semi-definite} matrix, but is not restricted to that. It extends the symmetric NMF by imposing a latent variable matrix $W$, so that it can allow matrix $B$ to provide a better approximation of clusters in practice \cite{ding2005equivalence}. A recent application of WS-NMF related to MIR can be found in musical structure segmentation \cite{Kauppinen2013}. 

\subsection{Vowel Subspace Learning with WS-NMF}
\label{ssec:vowel_subspace_wsnmf}
Next, we formulate the problem of the learning subspace that represents the repetitive vowel patterns from the input SSM with the following approximation:

\begin{equation}
\label{eq:wenmf2}
 S \approx B_{+}W B_{+}^{T},
\end{equation} 
where $S$ with $S \in \mathbb{R}^{N \times N}$, $B$ with $B \in \mathbb{R}^{N \times K}$, and $W$ with $W\in \mathbb{R}^{K \times K}$ represent the input SSM, unlabeled vowel activation indicator matrix, and transformation matrix of vowel subspace, respectively. Here, $N$ denotes a set of frame indices and $K$ denotes the number of vowel classes.

To gain a deeper understanding of Equation \ref{eq:wenmf2}, we can assume $S \in \mathbb{R}^{N \times N}$ to be factorized as $\beta \beta^{T}$ with $\beta \in \mathbb{R}^{K \times N}$.
At first glance, it is then of the same form that may be solved by various existing clustering methods, such as kernel K-means\cite{Dhillon:2004:KKS:1014052.1014118}, spectral clustering \cite{ng2002spectral}, and symmetric NMF. However, the SSM we obtained is non-negative and a mostly indefinite matrix. In this case, it is difficult to use $\beta$ as an indicator of vowel activation, due to the negative {\it eigenvalue} property of $S$. If we want a better representation of unlabeled vowel clusters, one possible explanation of Equation 8 can be reached by assuming $\beta$ as a multiplication of weight $\omega$ with matrix $B \in \mathbb{R}^{K,N}$.

\begin{equation}
\label{eq:wsnmf_illust}
S\approx \beta \beta^{T} = (\omega B^{T})^{T}(\omega B^{T}) = B(\omega^{T}\omega)B^{T}. 
\end{equation}

After rewriting this as the right-most part of Equation \ref{eq:wsnmf_illust}, it can be represented in the form of Equation \ref{eq:wenmf2}, if the transformation matrix W is a positive semi-definite matrix. Thus, the relaxation of Equation 8 imposing $W$ can provide more freedom for the expression of the subspaces associated with both positive and negative eigenvalues\cite{ding2005equivalence}. In this way, the non-negative matrix $B$ can be viewed as more a relevant observation of unlabeled vowel classes mapping over the matrix $\beta$.

\begin{algorithm}[t]
\label{algo:wsnmf}
\SetKwInOut{input}{Input}\SetKwInOut{output}{Output} 
\DontPrintSemicolon
\input {$S$, $K$ informed by $L'$}
\output {$B$, $W$}

\Begin{
$\colveccc[1]{\text{{\it Initialize }} B, W}$\;
\Repeat{$\colveccc[1]{\begin{Vmatrix}S - BWB^{T}\end{Vmatrix}_F^2 \text{converges.}}$ }{
\begin{align}
\label{eq:wsnmf_update1}
& W \leftarrow W \otimes \colveccc[.8]{
\dfrac{(W^{T}BW)}{(B^{T}BWB^{T}B + \epsilon)}
}
\\
\label{eq:wsnmf_update2}
& \widetilde{B} \leftarrow B \otimes
\colvecc[.8]{
.5+.5* \dfrac{(SBW)}{(BWB^{T}BW + \epsilon)}
}
\end{align}
\For{$\colvecc[.8]{\forall i, i\in N}$} 
{\begin{equation}B_{i} \leftarrow \colveccc[.8]{
\dfrac{-\widetilde{B}_{i}/\ell}{\Omega(-\widetilde{B}_{i}e^{1+\rho_i/\ell}/\ell)}
}
\end{equation}
}	
}
}
\SetAlgoLined
\caption{WS-NMF with sparsity}
\end{algorithm}

The update rules for $B$ and $W$ are given in Equation 10--11 of Algorithm 1, where $\otimes$ is the {\it Hadamard} product, $\div$ is an element-wise division of the upper term by the lower term, and $\epsilon$ is a small, positive constant. In practice, we initialize $B$ and $W$ with random values in the range between $10^{-4}$ and 1.

In Algorithm 1, it is important to design the unlabeled vowel activation indicator matrix $B$ and the transformation matrix $W$ with adequate length, $K$; however, note that $N$ is always fixed by the size of the given input SSM. Generally, we can expect that increasing $K$ would provide more discriminability of the model, while decreasing $K$ would simplify the model and lessen the processing time. In our case, the matrix $A$ oriented from the lyrics can provide a good hint for the choice of the ideal length for $K$. As a default, we set $K$ by counting $L'$, the number of vowel classes appearing in the lyrics of a target song, such that $L'\leq L$ with $L$ denotes the column length of $A$. Thus, we can define K as

\begin{equation}
\label{eq:get_k}
	K = L' + i, \quad s.t.~\text{integer~} i.
\end{equation}
Equation \ref{eq:get_k} allows us to de-/increase model complexity by adding an integer, $i$. Further discussion on the effect of adjusting $K$ will be presented in Section \ref{ssec:exp_WSNMF_K}.

Because a singer can make the sound of only one vowel at a time, enforcing $B_{\{i | i \in N\}}$ to have a sparse distribution can help in achieving more meaningful patterns. Here, we impose sparsity based on entropy, as detailed in \cite{shashanka2008sparse} over activations $B_{i}$. In Algorithm 1, Equation 12 constrains the sparsity for every iterative update, where $\Omega(\cdot)$ and $\rho$ denote {\it Lambert's function} and a {\it Lagrange multiplier}, respectively. In practice, the sparsity parameter $\ell = 3\times 10^{-3}$ was found to be effective.

So far, we have described the utility of matrix $W$ in terms of the matrix {\it eigenvalue} property, with a focus on its functionality in helping matrix $B$ to be more informative. Additionally, in practice, we could gain clues as to the nature of the vowel subspace---to give an example, a phonetics study claimed that the distance between the vowels /a/ and /i/ should be statistically larger than that between /e/ and /i/ \cite{obleser2003cortical}---in the output matrix $W$ that we obtained. This could lead to a possible semi-supervised scenario, such as fixing matrix $W$ with appropriate prior knowledge from previous phonetics studies. However, that idea is beyond the scope of the present research.

\section{Spatiotemporal Transformations and Alignment By Canonical Time Warping}
\label{sec:spatialtemporal_ctw}
Recall that we generated matrixes $A$ and $B$ in Sections \ref{sec:textProcessing} and \ref{sec:vowelSubspaceLearn}, respectively. The logical matrix $A$ represents a sequence of vowel repetition patterns observed in textual lyrics, while matrix $B$ represents vowel activations observed from acoustic features. Each matrix belongs to a different spatiotemporal space such that $A\in\mathbb{R}^{M\times L}$ and $B\in\mathbb{R}^{N\times K}$, where ${M} \neq {N}$ and ${L}\neq {K}$, respectively. In this regard, most of the existing methods for the alignment of DNA sequences\cite{lipman1985rapid}, audio frames\cite{sakoe1978dynamic}, and image pixels\cite{Szeliski:2006:IAS:1295184.1295185} are not readily applicable unless we assume at least one equivalent basis to be shared with, such that $\mathbb{R}^{M}=\mathbb{R}^{N}$. To overcome this, we used a spatiotemporal alignment method derived from the study of computer vision \cite{zhou2009canonical}. Additionally, we also considered several other applicable methods \cite{izumitani2008robust, martin2009musical,Hsu:2005:STH:1186822.1073315} as baselines in Section \ref{subsec:SA_in_KP}.

\subsection{Formulation of spatiotemporal transformation}

First, we define a temporal transformation required for aligning sequence $A$ to another sequence $B$. Assume that for $m = \{1,\ldots,M\}$, $\hat{A}_M$ is a time series. $\hat{A}_m \in \mathbb{R}^{\it d}$ is a {\it d}-dimensional vector, such that matrix $\hat{A} \in \mathbb{R}^{M \times {\it d}}$. Also, for $n=\{1,\ldots,N\}$, $\hat{B}_n \in \mathbb{R}^{\it d}$ is a {\it d}-dimensional vector, such that matrix $\hat{B} \in \mathbb{R}^{N \times {\it d}}$. In this way, $\hat{A}$ and $\hat{B}$ can be assumed to be spatially transformed matrixes of the original $A$ and $B$ into {\it d}-dimensions. Now, the temporal transformations for alignment between $\hat{A}$ and $\hat{B}$ in the target space $H$ can be represented as

\begin{equation}
	f_{\hat{A}}:M \rightarrow H, \quad f_{\hat{B}}:N \rightarrow H.
	\label{eq:temporal_trans}
\end{equation}

By defining the mapping matrixes $Q_a$ with $Q_a \in \mathbb{R}^{M \times H}$ and $Q_b$ with $Q_b \in \mathbb{R}^{N \times H}$, the alignment problem must be equivalent to a problem searching $\{Q_a,Q_b\}$ that minimizes

\begin{equation}
 \text{min}\begin{Vmatrix} {\hat{A}^T}{Q_a} - {\hat{B}^T}Q_{b}\end{Vmatrix}_F^{2}, 
 \label{eq:dtw}
\end {equation}
where ${\begin{Vmatrix} \cdot \end{Vmatrix}^2}$ denotes a {\it Frobenius norm}. If we define $Q$ as logical sequence matrixes (s.t. $q_a\in\{0,1\}$ and $q_b\in{0,1}$), then a well-known algorithm that solves Equation \ref{eq:dtw} by hard mapping is {\it dynamic time warping} (DTW)\cite{Szeliski:2006:IAS:1295184.1295185}.

Next, we define the spatial transformation required for mapping between labeled and unlabeled vowels, represented by the columns of $A$ and $B$, respectively. The transformations of the given original matrixes $A$ from $\mathbb{R}^{M \times L}$ to $\mathbb{R}^{M \times Z}$, and $B$ from $\mathbb{R}^{N \times K}$ to $\mathbb{R}^{N \times Z}$ are

\begin{equation}
	f_{A}:L \rightarrow Z, \quad f_{B}:K \rightarrow Z,	
\end{equation}
where the projective target space is denoted by $Z$. The purpose of this transformation is to project the two given matrixes into $d_Z$-dimensional space where the transformed matrixes can have maximum correlations for each dimension, one by one. {\it Canonical correlation analysis} (CCA) \cite{hardoon2004canonical} has provided a solution to this problem with a restricted sample size. By defining certain soft mapping matrixes $V_a$ with $V_a \in \mathbb{R}^{Z \times L }$ and $V_b$ with $V_b \in \mathbb{R}^{Z \times K }$, the objective of CCA is

\begin{equation}
 \text{min}\begin{Vmatrix} {{V_a}\bar{A}^T} - {V_{b}\bar{B}^T}\end{Vmatrix}_F^{2}. 
 \label{eq:cca}
\end {equation}

\begin{algorithm}[t]
\SetKwInOut{input}{Input}\SetKwInOut{output}{Output} 
\DontPrintSemicolon

\input {$A_{Lyric}, B_{Audio}$}
\output {$Q_{a}$, $Q_{b}$, $V_{a}$, $V_{b}$}\;
\Begin{
$\colveccc[1]{\text{\it Initialize } Q_{a}, Q_{b} \text{\it using UTW}(Q,h})$.\;\;
\Repeat{$\colveccc[1]{J_{ctw} \text{ converges.}}$}{
$V \leftarrow$ {\it Solve} $V=[V_{a}^{T}, V_{b}^{T}]^{T}$ {\it that obeys}:

\begin{equation} \label{eq:ccaSolution} \hspace{-.08cm}
\colvec[.8]{0 & A^{T}QB\\ B^{T}Q^{T}A^{T} & 0}V = \lambda 
\colvec[.8]{A^{T}D_{a}A & 0\\ 0 & B^{T}D_{b}B} V
\end{equation} 

\hspace*{.8cm} {\it using CCA, given $\varsigma$}.\;\;
$Q \leftarrow$ {\it Solve} $\begin{Bmatrix}Q_{a}, Q_{b}|V_a^{T}A^{T}, V_b^{T}B^{T}\end{Bmatrix}$ {\it using DTW}.\;
\;
}
}
\SetAlgoLined
\label{algo:ctw}
\caption{CTW for alignment}
\end{algorithm}

In Equation \ref{eq:cca}, both of the input matrixes $\{\bar{A},\bar{B}\}$ are required to have the same length of rows, due to the restriction of CCA that only allows finding relationships between two multivariable inputs with the same numbers of samples. This requirement should be satisfied with the temporal transformation above, such that $\bar{A} \leftarrow {\hat{A}^T} {Q_a}$ and $\bar{B} \leftarrow {\hat{B}^T} {Q_b}$. 

To summarize, the addressed transformations were:
\begin{itemize}
\item Temporal transformation, which finds the optimal alignment path between two different time series matrixes under restrictions of the input matrixes with the same lengths of columns, can be performed by DTW in hypothetical space $H$.
\item Spatial transformation finds a good mapping between the labeled vowel matrix and the unlabeled vowel activation matrix. It can be performed by CCA in the hypothetical space $Z$, only if the given input matrixes have the same lengths of rows.
\end{itemize}

\subsection{Alignment using CTW}
\label{ssec:alignment_using_ctw}
The rationale behind DTW and CCA sheds lights on an iterative process that uses the output of DTW as input for CCA, and the output of CCA as input for DTW. One such algorithm that extends DTW and CCA is canonical time warping (CTW) \cite{zhou2009canonical}, previously suggested for the task of human motion alignment in computer vision. To apply CTW to the alignment problem that we have addressed, the objective function can be written as

\begin{equation}
J_{ctw}(Q_{a},Q_{b},V_{a},V_{b}) = \begin{Vmatrix} {V_{a}}^{T}A^{T}Q_{a} - V_{b}^{T}B^{T}Q_{b} \end{Vmatrix}^{2}_{F},
\label{eq:jctw}
\end{equation}

where $A \in \mathbb{R}^{M \times L}$ and $B \in \mathbb{R}^{N \times K}$ are input matrixes with $\{M,N\}$ denoting the temporal basis and $\{L,K\}$ denoting the spatial basis, and 
$V_a \in \mathbb{R}^{Z \times L}$ and $Q_a \in \mathbb{R}^{M \times H}$ are spatiotemporal transformation matrixes for A; $V_b \in \mathbb{R}^{Z \times K}$ and $Q_b \in \mathbb{R}^{N \times H}$ are spatiotemporal transformation matrixes for B. Note that the temporal transformation matrix $Q$ is defined as $\{Q|q\in\{0,1\}\}$, for use with DTW. The total error $J_{ctw}$ can be estimated through the hypothetical space of $\mathbb{R}^{Z \times H}$. When $J_{ctw}$ converges, we take $Q$ as a result of the temporal alignment between the lyrics and audio. More specifically, the product of $Q_a$ and $Q_b$ represents the alignment path between $m= \{1,\ldots,M\}$th vowel of the lyrics and $n=\{1,\ldots,N\}$th frame of the audio features.

Algorithm 2 provides update rules that iteratively minimize Equation \ref{eq:jctw}. In each cycle, CCA searches optimal spatial transformation $V$ through the generalized {\it eigenvalue problem} ($Xy=\lambda y$), as shown in Equation \ref{eq:ccaSolution}, which provides a {\it closed form solution}\cite{hardoon2004canonical}. Then, the temporal transformation $Q$ is updated with the existing solution of DTW\cite{Szeliski:2006:IAS:1295184.1295185}. 

A remaining problem we have before addressing CTW is that at least one set of $Q$ or $V$ should be initialized appropriately to satisfy the accordance of sample size for CCA or the number of dimensions for DTW at the first iteration. 
Previously, Zhou et al.\cite{zhou2009canonical} suggested methods to initialize $V$ with identity matrixes or alternatively, {\it principal component analysis} (PCA) \cite{wold1987principal}. Another possible option we considered was to initialize $Q$ with uniform time warping (UTW) \cite{Fu:2008:STW:1380759.1380764}, which increases the sample rate of the input matrixes $\{A,B\}$, so that the sample sizes of $A^{T}Q_A$ and $B^{T}Q_B$ can be equal. UTW initializes $Q_A$ and $Q_B$ to be matrixes with ones on the diagonal, and the entries outside are all zeros. Through preliminary tests, we found that initialization by UTW was more stable than other initialization methods.

In Algorithm 2, it is important to design the transformation matrix $\{Q,V \}$ with an adequate size of target dimension. In the temporal domain, projection to higher dimensions, such that $H>\varrho \cdot \text{round}(\text{max}(M,N))$ with $\varrho > 1$, generally encourages achieving higher timing accuracy with smaller intervals in alignment results. In that way, the timing accuracy can be controlled by setting the length of rows $h$ for UTW in the initialization. In the spatial domain, CCA fundamentally performs a dimensional-reduction function. Thus, the column length of $Z$ must obey $1 \leq Z\leq \text{min}(L,K)$. Obviously, there will be a trade-off between avoiding  oversensitivity and an oversimplified model. We use an existing implementation of CCA\cite{hardoon2004canonical}. It provides a hyperparameter $\varsigma$ that controls $Z$ with regard to the amount of energy kept.

\section{Evaluation Methods}
\label{sec:evaluation}

\begin{table}[ht]
\caption{Design of Experiments}

\begin{threeparttable}

		\begin{tabular}{m{1cm} m{1cm} m{1.5cm} m{3.6cm} } 
    \toprule
		\textbf{Name} & \textbf{Dataset} & \textbf{Evaluation Metric} & \textbf{Purpose}\\
		\midrule
		\ref{subsec:SA_in_KP} & K-pop (KP) & Syllable-level Accuracy (SA) & A main experiment that measured syllable-level accuracy of the present system. Due to the lack of an existing data set with syllable-level annotation, it was necessary to construct KP and define the SA metric. The result was compared with multiple baseline methods.\\
		\hline
		\ref{subsec:WA_in_EP} & English-pop (EP) & Word-level Accuracy (WA) & A main experiment that evaluated our system at word-level accuracy, and compared the performance with the existing ASR-based supervised system. Using EP and WA, we kept equivalent experimental conditions to those described in the previous work\cite{mauch2012integrating}.\\
		\hline
		\ref{ssec:Validation_vowel_only} & KP & SA & Validation of our prior assumption, `Observing only vowels', by examining theoretical upper-bound of accuracy.\\
		\hline
		\ref{ssec:exp_WSNMF_K} & KP & SA & Tuning hyperparameters of WS-NMF+CTW by a grid-search method.\\
		\hline
		\ref{ssec:effect_of_VAD} & KP, {\it Jamendo}\tnote{$\dagger$}& SA & 1) VAD benchmark test, 2) Effect of VAD error.\\
		\bottomrule

		\end{tabular}
	\begin{tablenotes}
      \item[$\dagger$] Used for training and validation of a baseline VAD algorithm.
    
  \end{tablenotes}		
\end{threeparttable}

\label{table:experiment_cat}
\end{table}

The proposed process was evaluated in two major experiments (\ref{subsec:SA_in_KP} to \ref{subsec:WA_in_EP}), and additional experiments (\ref{ssec:Validation_vowel_only} to \ref{ssec:effect_of_VAD}) within the specified data sets, as shown in Table \ref{table:experiment_cat}.

\subsection{Dataset description}
\label{ssec:dataset}
\subsubsection{K-pop (KP)}

The data set contained 11 popular Korean music recordings published within the last 3 years by five male and six female solo singers. In Table \ref{table:songs_in_KP}, a set of the last three songs was used as a validation set for tuning hyperparameters in Section \ref{ssec:exp_WSNMF_K}. With respect to the genre distribution in the test set, 37.50\% of the songs were categorized as dance, whereas the other 62.50\% were ballads. The audio files were produced in stereo with a sampling rate of 44,100 Hz. The syllable-level lyric annotations with onset and offset times were cross-validated by three persons who spoke Korean as their primary language. The consonants are not annotated as a separate component in the syllables. Thus, an annotated pair of onset and offset indicates the actual margin of one sung syllable that typically includes an initial consonant and its combined vowel, and its final consonant, if available.

\begin{table}[t]
\caption{Songs in the K-pop data set}

\begin{threeparttable}

		\begin{tabular}{m{2cm} m{4cm} m{1.4cm} } 
    \toprule
		\textbf{Artist} &\textbf{Title} & \textbf{Category\tnote{$\ddagger$}}\\
		\midrule
		Younha & Not there & BF\\
		G-Dragon & Crooked & DM\\
		Jung Jinu & Satellite & DM\\
		Hyorin & Let it go (from ``Frozen'', Korean OST) & BF\\
		Seung Ree & Gotta talk to U & DM\\
	
		Ahn Ye Eun & Sticker & BF\\
		Jung Seun & Mom, just a moment & BM\\
	
		Woo Yerin & Fish Bowl & BF\\
		AOA & Miniskirt & DF-V\\
		Lee Seol Ah & Life as a mom & BF-V\\
		Busker Busker & Cherry blossom ending & BM-V\\
		\bottomrule

		\end{tabular}	
	\begin{tablenotes}
    \item[$\dagger$] B: ballad, D: dance, F: female, M: male, V: held-out validation set.
    
  \end{tablenotes}	
\end{threeparttable}

\label{table:songs_in_KP}
\end{table}

\subsubsection{English-pop (EP)}
To compare our system with an existing ASR-based system for {\it lyrics-to-audio} alignment, we obtained a database from the authors\cite{mauch2012integrating}. The data set contained $17$ uncut original versions of international pop songs, mostly published in the 1980--90s. The rest were a cover song and two songs from the RWC Music Database\cite{goto2002rwc}. These songs were provided with ground-truth word onsets manually annotated by the author's group. The duration of each song varied in the range from 3 to 6 min. The audio files were produced in stereo with a sampling rate of 44,100 Hz. In total, the database contained 5,052 words within 4,753 s.

\subsubsection{{\it Jamendo}}
In Section \ref{ssec:effect_of_VAD}, we used the {\it Jamendo}\cite{jamendo_1998} data set for training, parameter optimization, and testing of the baseline VAD algorithm, based on Lehner\cite{lehner2013towards}. This data set consisted of 61 songs for training, 16 for validation, and 16 for testing. The audio files were produced in stereo with a sampling rate of 44,100 Hz. The songs were annotated manually in two classes, for voice-active or -inactive.

\subsection{Evaluation Metrics}
\label{ssec:Evaluation_Metrics}
 
\subsubsection{Syllable-level Accuracy (SA)}
\label{sssec:SyllableAcc}
This metric inherits most of practical properties from the word-level evaluation metric introduced in the related report \cite{mauch2012integrating}. We extended the metric for use in a syllable-level evaluation of the KP data set. We assumed the number of songs in the test data set to be $\Gamma_{\text{songs}}$, and the number of syllables $v$ in a song $s$ to be $\Gamma_{v|s}$. From this, the mean accuracy over the songs in the data set can be defined as 

\begin{equation}
\label{eq:SA}
	\text{SA(\%)} = \dfrac{1}{\Gamma_{\text{songs}}}
	\sum_{\text{song }s}
		\underbrace{
			{\left\{{ \dfrac{1}{\Gamma_{v|s}} \sum_{\text{syllable }v} 
			1_{|t_v-\hat{t}_v|<\tau}
			}\right\}}
			\cdot 100
		}_{=~\Theta_{\tau}(s), \text{~mean accuracy of }s\text{th song}},
\end{equation}
where $t_v$ and $\hat{t}_v$ are the true and predicted alignment times for the syllable $v$, respectively. Here, the indicator function $1_{(\cdot)}$ outputs 1 if the absolute alignment error $|t_v-\hat{t}_v|$ is smaller than tolerance $\tau$ ($= 1$ second as a default), otherwise, zero. By controlling the tolerance value, this metric can provide a practical measure of alignment quality, meeting various purposes for use in a real-world system.

\subsubsection{Word-level Accuracy (WA)}	
This metric was presented in previous work \cite{mauch2012integrating} with their data set, EP. The mean accuracy over the number of songs, $\Gamma_{\text{songs}}$, was defined as

\begin{equation}
\label{eq:WA}
	\text{WA(\%)} = \dfrac{1}{\Gamma_{\text{songs}}}
	\sum_{\text{song }s}
		\underbrace{
			{\left\{{ \dfrac{1}{\Gamma_{w|s}} \sum_{\text{word }w} 
			1_{|t_w-\hat{t}_w|<\tau}
			}\right\}}
			\cdot 100
	}_{=~\Theta_{\tau}(s), \text{~mean accuracy of }s\text{th song}},
\end{equation}
where $\Gamma_{w|s}$ denotes the number of words $w$, given a song $s$. Here, $t_w$ and $\hat{t}_w$ denote the true and predicted alignment times for the word $w$, respectively.

\subsubsection{Standard Deviation (STD) of Accuracy}
We estimated the unbiased variance of SA or WA, among songs in a data set as

\begin{equation}
	\text{STD}_{\tau} = \sqrt{\dfrac{1}{\Gamma_{\text{songs}}-1}
		\sum_{\text{song~}s}{(\Theta_{\tau}(s) - \mu(\Theta_{\tau}))^2}
	},
\end{equation}
where $\Theta_{\tau}(s)$ is song-wise mean accuracy, defined in Equation \ref{eq:SA} at the syllable-level, or Equation \ref{eq:WA} at the word-level.

\subsubsection{Mean Absolute Deviation (MAD)}
We measured the mean absolute deviation of alignment error in seconds over songs by

\begin{equation}
\label{eq:MAD_vw}
	\text{MAD}_{\pi} = \dfrac{1}{\Gamma_{\text{songs}}}
	\sum_{\text{song s}}	
	{\left\{
		\dfrac{1}{\Gamma_{\pi}}
		\sum_{\text{syllable or word,}~\pi} 
		{|\hat{t}_\pi - t_\pi|}
	\right\}},
\end{equation}
where $\pi$ should be replaced by $v$ or $w$, representing the syllable- or word-level estimation of MAD, respectively.

\subsubsection{Metrics for Evaluation of VAD}
The binary result of VAD described in Section \ref{ssec:effect_of_VAD} can be evaluated in terms of the precision ($P$), recall ($R$), and F-measure. Because we only took the positive (predicted to be voice-active) frames as an input for the main algorithm for alignment, we focused on the count of true-positives($tp$), false-positives($fp$), and false-negatives($fn$), rather than true-negatives($tn$).
We defined $P={tp}/({tp+fp})$, $R={tp}/({tp+fn})$, and $F_1 = 2 \cdot ({P \cdot R})/({P + R})$. We also made use of the $F_{\beta}$-measure to assess the tradeoff between precision and recall of VAD in the proposed system.
Generally, $F_2$ weights recall higher than precision, while $F_{0.5}$ gives more emphasis to precision than recall.

\subsection{Evaluation Setup}
The five experiments presented in the Results tested and compared the proposed method with various other methods, by taking a pair of uniformly processed inputs. The pairs of inputs are denoted as $\text{Input}_{\text{dataset}}$, such that $\{A_{\text{dataset}}, S_{\text{dataset}}\} \in \text{Input}_{\text{dataset}}$, where $\{\text{KP,EP}\}\in \text{dataset}$. Also, $\{A_{\text{song}},S_{\text{song}}\}$ denotes a pair of the inputs that are involved in a song from the target data set. Input $A_{\text{dataset}}$ was obtained by passing the lyrics part of the data set through the text processor with the parameters suggested in Section \ref{sec:textProcessing}. Input $S_{\text{dataset}}$ was a SSM obtained by passing the audio part of the data set through the chains of pre-processors (singing voice separation$\rightarrow$ VAD $\rightarrow$ feature extraction $\rightarrow$ Generation of SSM) with the parameters suggested in Section \ref{subsec:Pre-processing_audio}.

Note that the proposed method is denoted as {\bf WS-NMF+CTW} in the Results. The results came from the chain of WS-NMF$\rightarrow$CTW as follows:

\begin{enumerate}
	\item We first passed the set of input $\{ A_{\text{dataset}}, S_{\text{dataset}} \}$ through WS-NMF. We uniformly stopped Algorithm 1 after 5,000 iterations. The output of WS-NMF was denoted as $B_{\text{dataset}}$.\\    
	\item Next, we passed the above result, $\{ A_{\text{dataset}}, B_{\text{dataset}} \}$ through CTW. The hyperparameter $\varsigma$ was set to keep 95\% (default) of the energy in CCA. In most cases, Algorithm 2 stopped within 25 cycles, according to the stopping criterion, $|J_{\text{ctw}}(t)-J_{\text{ctw}}(t-1)| < 10^{-4}$. 
	\item Finally, the alignment result for each song was obtained by multiplying the output of CCA, such that $\{Q_a \cdot Q_b\}_{\text{song}}$. We evaluated the average performance in the data set using the metrics given in Section \ref{ssec:Evaluation_Metrics}.
\end{enumerate}

Since the proposed WS-NMF+CTW had adjustable hyperparameters such as $K$, $\varsigma$, and $\ell$, we performed hyperparameter-tuning as detailed in \ref{ssec:exp_WSNMF_K}. Then, we applied the tuned hyperparameters for generating the main results reported in  \ref{subsec:SA_in_KP} and \ref{subsec:WA_in_EP}.

\section{Results}
\label{sec:result}
\begin{table}[htbp]
\caption{Description of Baseline Algorithms in \ref{subsec:SA_in_KP}}
\centering
		{
		\begin{tabular}{m{0.7cm} m{1.2cm} m{5.8cm} } 
    \toprule
		\textbf{Name} & \textbf{Input} & \textbf{Description}\\
		\midrule

		{IK} &$\{ A^{\text{SSM}}_{\text{song}}$ $, S_{\text{song}} \}$ & Izumitani \& Kashino\cite{izumitani2008robust} proposed a 2-$D$ DTW-based method for key-invariant audio matching. Later, their method was shown to be applicable to an alignment of correlated patterns in two input SSMs\cite{martin2009musical}. However, it has not been tested before in SSMs as we did. Because $A_{\text{song}}$ of our given set of input $\{A_{\text{dataset}}$ $, S_{\text{dataset}}\}$ was not a SSM, we first generated a binary SSM (let us denote, $A^{\text{SSM}}$) from $A_{\text{song}}$ by calculating the {\it Hamming distance}. Then, IK was performed using $\{ A^{\text{SSM}}_{\text{song}}, S_{\text{song}} \}$ as a song-wise input.\\
		\hline
		{Martin} &$\{ A^{\text{SSM}}_{\text{song}}$ $, S_{\text{song}} \}$ & As an extension of IK, Martin, {\it et al.}\cite{martin2009musical} proposed to replace the cost function with a hand-crafted one---namely, {\it adapted Euclidean distance}\cite{martin2009musical}---for the task of musical structure retrieval.\\ 
		\hline
		{CTW} & $\{A_{\text{song}}$ $, B_{\text{MFCCs}}\} $ & Without the help of the proposed combination with WS-NMF, CTW\cite{zhou2009canonical} can work as a stand-alone alignment machine, by taking $\{ A_{\text{song}}, B_{\text{MFCCs}} \}$ as a set of inputs for each song. Here, $B_{\text{MFCCs}}$ represents the scaled feature with zero mean and unit variance normalization of the MFCCs feature. \\
		\hline
		{pCTW} & $\{A_{\text{song}}$ $, B_{\text{MFCCs}}\} $& As an extension of CTW, pCTW uses the PCA derivatives of the original features. \\
		\hline
		{GTW} & $\{A_{\text{song}}$ $, B_{\text{MFCCs}}\} $& Generalized time warping (GTW) \cite{zhou2012generalized} extended CTW by parameterizing the temporal warping as a combination of monotonic basis functions and improved optimization. We used the implementation of the authors\cite{zhou2012generalized}, and tested a recommended regularization weight, $\xi=1$. \\

		\bottomrule
		
		\end{tabular}
		}
\label{table:baselines}
\end{table}

\subsection{Syllable-level Accuracy (SA) in KP}
\label{subsec:SA_in_KP}

\begin{table*}[!t]
\caption{Main Result of \ref{subsec:SA_in_KP}: Syllable-level Accuracy (\%) and Mean Absolute Deviation ({\it s}) in the KP data set, with VAD=Predicted ($P=72.58\%, R=93.07\%, {F}_1 =81.56\%$).}
\centering
		{
		\begin{tabular}{p{1.9cm} p{2cm} p{2cm} p{2cm} p{2cm} p{2cm} p{2.3cm} } 

			\toprule
			 &\bf{IK} & \bf{Martin} & \bf{ CTW }& \bf{pCTW} & \bf{GTW} & \bf{WS-NMF+ CTW}\\	
			
			\midrule
			
			\bf{SA$\pm\text{STD}_{\tau=1s}$} & 13.15$\pm10.42$ & 11.70$\pm6.43$ & 26.31$\pm14.78$ & 26.64$\pm07.72$ & 26.51$\pm07.09$ & 62.76$\pm12.82$\\
		
			\bf{$\text{MAD}_v$} & 28.70 & 31.99 & 15.65 & 10.68 & 12.83  &3.04\\
			\bottomrule 
		\end{tabular}
		}
\label{table:result_KP_main}
\end{table*}

\begin{table*}[!t]
\caption{Sub-Result of \ref{subsec:SA_in_KP}: Syllable-level Accuracy (\%) and Mean Absolute Deviation ({\it s}) in the KP data set, with $\text{VAD=True}$. }
\centering
		{
		\begin{tabular}{p{1.9cm} p{2cm} p{2cm} p{2cm} p{2cm} p{2cm} p{2.3cm} } 
			\toprule
			 & \bf{IK} & \bf{Martin} & \bf{CTW }& \bf{pCTW} & \bf{GTW} & \bf{WS-NMF+ CTW}\\	
			\midrule 
			\bf{SA}$\pm\text{STD}_{\tau=1s}$ & 16.11$\pm10.27$ & 15.60$\pm11.51$ & 33.40$\pm29.06$ & 41.02$\pm13.40$ & 41.34$\pm11.06 $&  69.34$\pm12.07$\\
		
			\bf{$\text{MAD}_v$} & 19.27 & 20.82 & 7.23 & 4.57 & 4.58  &1.94\\
	
			\bottomrule
		\end{tabular}
		}
\label{table:result_KP_sub}
\end{table*}

We evaluated the proposed system by measuring the alignment accuracy at syllable level, averaged over all the songs in the KP data set. Although we could not find any previous work dealing with {\it lyrics-to-audio} alignment in a completely unsupervised manner, several existing algorithms that were developed for the general purpose of spatiotemporal alignment could be regarded as potential candidates. In Table \ref{table:baselines}, we reviewed such algorithms as multiple baselines.

\subsubsection{Performance of the proposed system}
\label{ssec:Performance_KP}
Table \ref{table:result_KP_main} displays the main result. The values in the first row indicate that the entire systems, consisting of the same pre-processor paired with different alignment algorithms, produced the percentile of correct alignments within the $\tau=$1 s tolerance of temporal displacement. Note that the pre-processor included the predicted VAD with the suggested threshold parameter $\theta = 1.88$. The values after $\pm$ represent the unbiased variance of SA that came from song-wise evaluations. WS-NMF+CTW marked the highest SA of $62.76\%$, while the second highest SA of $26.64\%$ was barely achieved by pCTW. Overall, WS-NMF+CTW outperformed the most recent multimodal alignment algorithms, such as CTW, pCTW, and GTW, based on unsupervised learning. The values of MAD in the second row revealed the mean alignment displacement error in time. WS-NMF+CTW showed the lowest MAD, of 3.04 s, among the algorithms compared.

\subsubsection{Performance of VAD in the Preprocessors}
\label{ssec:Performance_of_VAD_KP}
We evaluated the binary classification accuracy of the predicted VAD independently, within the chain of preprocessors. The {\it precision} ($P$), {\it recall} ($R$), and $F_1$-measure were 72.58, 93.07, and 81.56\%, respectively. For more details, see Section \ref{ssec:effect_of_VAD}.

\subsubsection{Performance with True VAD}
\label{ssec:Performance_excluding_VAD_KP}
The sub-results displayed in Table \ref{table:result_KP_sub} were measured by replacing the VAD component in the preprocessor with ground truth onsets and offsets, which were available in the KP data set. However, the rest of the procedures were the same as in the main experiment above. This enabled us to test the proposed and baseline algorithms by excluding any external effects of VAD.

\begin{itemize}
	\item Given the true VAD, WS-NMF+CTW marked the highest SA of 69.34\% and the lowest MAD of 1.94 s, strongly outperforming all other baseline systems. In the comparison in Table \ref{table:result_KP_main} and \ref{table:result_KP_sub}, the SA for WS-NMF+CTW improved by 6.58\% points of a total 35.24\% points error, after replacing the predicted VAD with the ground truth. If the better VAD method was used, the proposed WS-NMF+CTW would be expected to produce better alignment performance, and the SA would reach near the upper bound of 69.34\%.
	\item The IK and Martin systems achieved the lowest SA and the longest MAD, in both Table \ref{table:result_KP_main} and \ref{table:result_KP_sub}. These results consistently showed that the two previous methods, originally proposed for much simpler pattern alignments (i.e., A-B-A-C-A for typical annotations of a certain music structure\cite{martin2009musical}), were not useful for our task.
	\item A sharp increase in SA for the three baseline algorithms---CTW ($\Uparrow$26.95\% for 26.31\%$\rightarrow$33.40\%), pCTW ($\Uparrow$53.98\% for 26.64\%$\rightarrow$41.02\%) and GTW ($\Uparrow$55.94\% for 26.51\%$\rightarrow$41.34\%)---was observed in comparing Table \ref{table:result_KP_main} to \ref{table:result_KP_sub}. This suggested that the existing multimodal alignment algorithms suffered from severe bottleneck problems from the predicted VAD. In contrast, the proposed algorithm ($\Uparrow$10.48\% for 62.76\%$\rightarrow$69.34\%) was relatively robust against VAD error versus the baseline algorithms.
\end{itemize}

\subsubsection{Alignment error deviations}

\begin{figure}[t]
	\centerline{{
	\includegraphics[width=1\columnwidth]{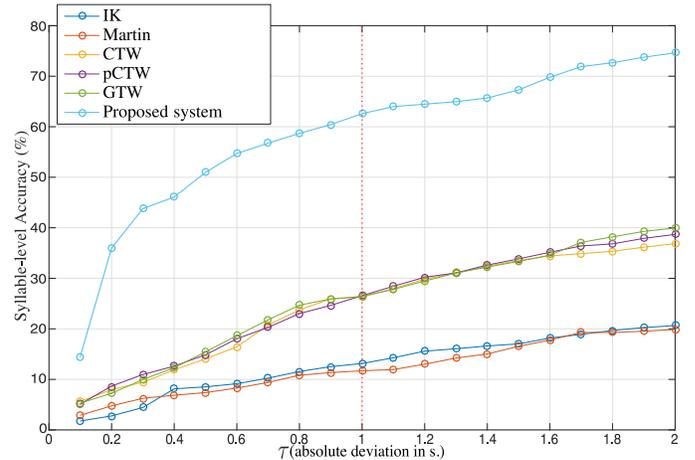}}}
	\caption{Syllable-level accuracy (SA) in the KP data set according to different values of $\tau$, in the range from 0.1 to 2.0 s. The red dotted vertical line marks the accuracy with $\tau=$1s, which we presented in the Results in Table \ref{table:result_KP_main}.}
 	\label{fig:KP_SA_tau}
\end{figure}

Figure \ref{fig:KP_SA_tau} shows the SAs measured by defining different tolerances, $\tau$. For the proposed system, the steepest increase in SA in the range between 0.1 and 0.3 s showed that the largest proportion of the error was distributed in the corresponding range. Given the most generous tolerance of $\tau=$ 2.0 s, the SA of the present system reached a maximum value of 75.58\%.

\subsection{Comparison with an ASR-based system: WA in EP}
\label{subsec:WA_in_EP}

In Table \ref{table:result_EP}, we compared the WA of the proposed system with the state-of-art ASR-based system\cite{fujihara2011lyricsynchronizer}. Our system consisted of the preprocessor, text processor, and WS-NMF+CTW, as described above. Because the compared ASR-based system\cite{fujihara2011lyricsynchronizer} was only evaluated at the word level, we followed the evaluation metrics provided in the compared work\cite{mauch2012integrating}, and used annotations for the EP data set obtained from the authors\cite{mauch2012integrating}.

\begin{table}[t]
\caption{Result of \ref{subsec:WA_in_EP}: Performance Comparison of the Two Systems in terms of Word-level Accuracy (\%) and Mean Absolute Deviation ({\it s}) in the EP data set\cite{mauch2012integrating}.}

\centering
{
	\begin{threeparttable}
		\begin{tabular}{p{2cm} p{1.8cm} p{1.8cm} } 
			\toprule
			 & \textbf{ASR-based,\newline Supervised\cite{fujihara2011lyricsynchronizer}} & \textbf{Proposed,\newline Unsupervised}\\	
			\midrule 
			WA$\pm\text{STD}_{\tau=1}$(\%) & 46.4$\pm22.3$\tnote{$\dagger$} & 61.32$\pm14.28$ \\

			$\text{MAD}_{\text{w}}$ (s) & 4.9\tnote{$\dagger$} & 2.92 \\
			\bottomrule 
		\end{tabular}
		
	\begin{tablenotes}
      \item[$\dagger$]The values were obtained from Mauch, {\it et al.}\cite{mauch2012integrating}, testing the original implementation of the ASR-based system\cite{fujihara2011lyricsynchronizer} in EP.
	\end{tablenotes}
	
	\end{threeparttable}
}

\label{table:result_EP}
\end{table}

\begin{figure}[ht]
	\centerline{{
	\includegraphics[width=1\columnwidth]{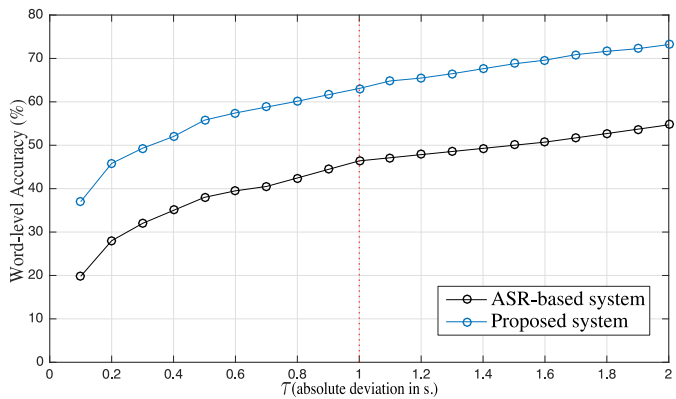}}}
	\caption{Word-level accuracy (WA) in the EP data set according to different values of $\tau$, in the range from 0.1 to 2.0 s. The red dotted vertical line marks the accuracy with $\tau=$1.0 s, which we presented in the Results in Table \ref{table:result_EP}. Note that the WA for the reported ASR-based system\cite{fujihara2011lyricsynchronizer} is  from the graph presented in the paper\cite{mauch2012integrating}.}
 	\label{fig:EP_WA_tau}
\end{figure}

The values in the first row of Table \ref{table:result_EP} indicate that the two systems compared produced the average percentile of correct alignments within the $\tau=$ 1.0 s tolerance of temporal displacement. The values after $\pm$ are the unbiased variance of WA that came from song-wise evaluations. WS-NMF+CTW showed a WA of 61.32\%, which was 32.16\% (14.92\% points) higher than the ASR-based system. The proposed system’s relatively lower standard deviation of 14.28\%, in comparison with ASR-based system's 22.3\%, indicated that the achieved alignment quality was more stable over the songs. The proposed system had a MAD of 2.85 s, while ASR-based system's value was 4.9 s. This lower MAD indicated that the proposed system produced 42.6\% point shorter word displacement errors, on average. Overall, the WA of the present system consistently outperformed the system compared in evaluations with various $\tau$ settings, as shown in Figure \ref{fig:EP_WA_tau}.

\subsection{Validation of the `Vowel-only' Assumption}
\label{ssec:Validation_vowel_only}

In Section \ref{ssec:Assumption}, we made the assumption that {\it lyrics-to-audio} alignment should be effectively achieved by `observing only vowels', while ignoring consonant parts of syllables. Because this assumption is a significant aspect of the present system, we needed a preliminary test to confirm its validity with real data. As a way to independently estimate the theoretical boundary of SA with this assumption, we used a set of $\{A_{\text{dataset}},S^{\text{g.t.}}_{\text{dataset}} \}$ as an input to DTW. The matrix $S^{\text{g.t.}}_{\text{dataset}}$ was generated directly using the annotated ground-truth vowel labels with onset locations in KP\footnote{Note that we could not test our prior assumption directly with the EP data set, where only word-level annotations with onsets were available.}. Thus, this input could be regarded as the perfect retrieval of vowels from audio features. Because the length of the input song may be another interesting issue in this test, we also simulated various conditions with respect to input length. We could generate them by random cuts of locally complete and at-least-five-syllables-activated original data in KP.

\begin{table}[t]
\caption{Results of \ref{ssec:Validation_vowel_only}: Theoretical Upper-Bound of Syllable-level Accuracy (SA) with the Assumption of Complete `Vowel-Only' Retrieval in KP.}
\centering
		{
		\begin{tabular}{p{2.1cm} p{1.1cm} p{1.1cm} p{1.1cm} } 
			\toprule
			\textbf{Input} &\textbf{VAD}& \textbf{SA(\%), $\tau=1$s} & $\textbf{MAD}_{v}$(s)\\	
			\midrule
			KP (no-cut)& True & 97.89 & 0.11 \\
			\midrule
			Random 30s cut & True & 97.84& 0.13 \\
			Random 20s cut & True & 97.93 & 0.18 \\
			Random 10s cut & True & 95.11 & 0.37 \\
	
			\midrule
			KP (no-cut) & Predicted & 91.50 & 0.71 \\
			
			\bottomrule
		\end{tabular}
		}
		
\label{table:result_vowel_only}
\end{table}

In the top row of Table \ref{table:result_vowel_only}, the theoretical upper-bound of SA in the KP data set was 97.89\%. Although one could provide an exceptional counterexample---a few existing songs repeat entirely {\it `la la la,...'} in the lyrics---, the experimental results caused us to accept the validity of the assumption as a near truth.

The three rows in the middle of Table \ref{table:result_vowel_only} present simulation results with three different shorter input lengths from an average of 20 tests using random cuts. The SAs were maintained, consistently higher than 95\%. Interestingly, MAD increased in a manner inversely proportional to the length of the input. Previously, the result with KP (no-cut) reached the highest SA and the lowest MAD, in comparison with the shorter random cuts. With the limitations of our results, this revealed that the longer input length would generally produce better alignment quality, assuming that a perfect VAD was provided.

The bottom row of Table \ref{table:result_vowel_only} shows the alignment results with the {\it ideal} features and the predicted VAD ($F_1$=81.56\%, evaluated in Section \ref{ssec:effect_of_VAD}). The resulting SA was 91.50\%, with a MAD of 0.71 s. This can be interpreted as the theoretical upper bound of SA with the proposed system in the KP data set.

\subsection{Hyperparameters of WS-NMF+CTW}
\label{ssec:exp_WSNMF_K}

\begin{figure}[!b]
	\centerline{{
	\includegraphics[width=1\columnwidth]{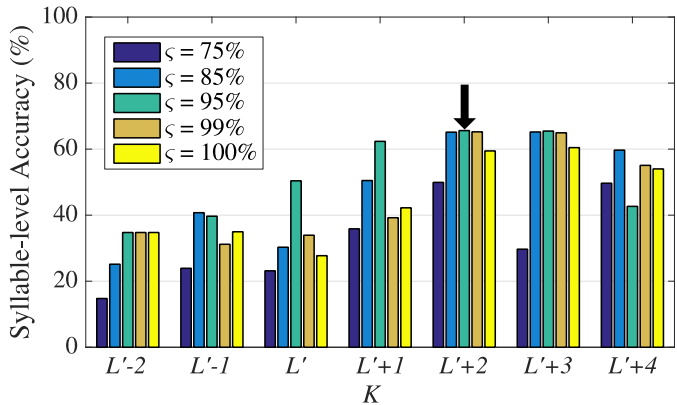}}}
	\caption{Syllable-level accuracy obtained by various settings of hyperparameters $\{K, \varsigma\}$: For $K$, $K=L'$ is the default value, informed by the number of vowel classes in the text. Each group represents the setting of $K$. In each group, five settings of $\varsigma$ were tested. The arrow points towards the optimum value.}
 	\label{fig:KP_opt_K}
\end{figure}

In advance of the main experiment \ref{subsec:SA_in_KP}, we tuned the hyperparameters for WS-NMF+CTW. In Section \ref{ssec:vowel_subspace_wsnmf}, Algorithm 1 for WS-NMF had two hyperparameters: $\ell$ for sparsity and $K$ (the number of clusters) for fixing the column length of the matrix $B$. In Section \ref{ssec:alignment_using_ctw}, Algorithm 2 for CTW had one hyperparameter, $\varsigma$ (\%), forcing the percentile of energy to be kept in the CCA feature vectors.

Among these three hyperparameters, we first found that $\ell = 3\times 10^{-3}$ generated stable results in preliminary tests. With respect to $K$ and $\varsigma$, generally, a larger value of $K$ increased model complexity and a larger $\varsigma$ decreased the complexity by dimensional reduction. In this regard, we tested various settings of $K$ with $\varsigma$, while $\ell$ was fixed.

For the grid search, we used the held-out validation set of KP in Table \ref{table:songs_in_KP}. The default value for $K$ was set by Equation \ref{eq:get_k} in Section \ref{ssec:vowel_subspace_wsnmf}, where $L'$ is the number of vowel classes appearing in the lyrical text, and the controllable $i$. We tested seven different settings for $K$ with $i\in\{-2,-1,0,1,2,3,4\}$, where $i=0$ was a default. For $\varsigma$, we tested five settings with $\varsigma (\%)\in\{72,88,95,99,100\}$. In combination, we tested 35 settings in total.

The results are shown in Figure \ref{fig:KP_opt_K}. Overall, we could observe that the settings for higher model complexity with $K>L'$ produced better results than the default setting and other settings for lower model complexity, such that $K\leq L'$. Then, the arrow in Figure \ref{fig:KP_opt_K} points to the best SA of 65.24\%, which was obtained setting $K=L'+2$ and $\varsigma=95\%$.

\subsection{VAD performance and its effects on the proposed system}
\label{ssec:effect_of_VAD}
We performed a benchmark test of the proposed VAD preprocessor described in Section \ref{ssec:VAD_RPCA}. We then investigated the effects of VAD false positives ($fp$) and false negatives ($fn$). 

\begin{figure}[t]
\begin{minipage}[t]{0.5\linewidth}
  \includegraphics[width=\linewidth]{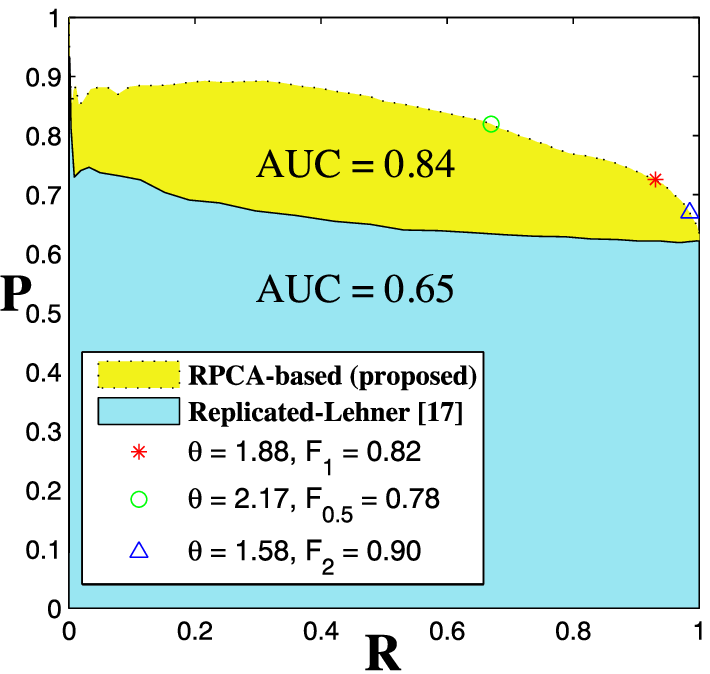}
  \caption{PR-curve of the proposed VAD, in comparison with the implementation of Lehner\cite{lehner2013towards} tested with the KP data set. The three points marked with *, $\circ$, and $\triangle$ represent the positions achieving the maximum $F_1$, $F_{0.5}$, and $F_2$ points, respectively.}
  \label{fig:pr_curve}
\end{minipage}%
  \hfill%
\begin{minipage}[t]{0.47\linewidth}
  \includegraphics[width=\linewidth]{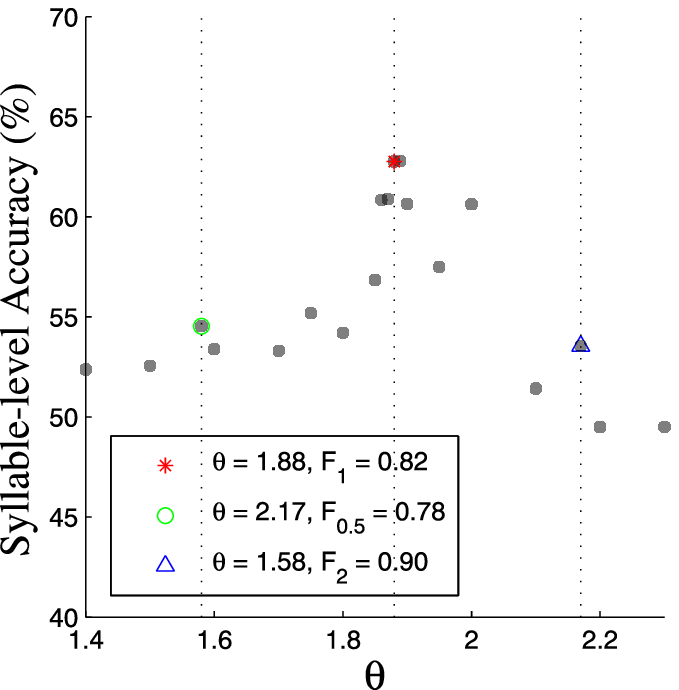}
  \caption{Sampled response graph of syllable accuracy (SA) in KP, according to the threshold parameter ($\theta$) of the proposed VAD. The best SA was obtained with $\theta=1.88$ around the position achieving the maximum $F_1$-measure for VAD.}
  \label{fig:thr_SA_response}
\end{minipage} 
\end{figure}


\subsubsection{Baseline Algorithm}
We implemented a baseline VAD algorithm based on the hand-crafted features and supervised learning previously proposed by Lehner\cite{lehner2013towards}. We extracted the audio feature (60 dimensions including $\delta$-MFCCs) with a total window of 800 ms around the 200 ms center frame, from the 22,050 Hz down-sampled input audio. A random forest\cite{breiman2001random} classifier was then trained with 61 songs, and parameter-tuned with another 16 unused songs included in the {\it Jamendo}\cite{jamendo_1998} data set. After performing postprocessing as suggested in the paper\cite{lehner2013towards}, the baseline implementation achieved $F_1$=81.90\% for the 16 test songs included in {\it Jamendo}. The performance of our implementation was 2.70\% points lower than in Lehner's report\cite{lehner2013towards}. 

\subsubsection{VAD Performance Benchmark}
In Figure \ref{fig:pr_curve}, we compared the present VAD with the baseline in the KP data set. The larger area under the curve (AUC) showed that the present VAD outperformed in terms of both {\it precision} and {\it recall}. The maximum $F_1$-measure achieved with the present VAD was 81.56\% (at $\theta=1.88$), which was 17.04\% points higher than the baseline result of $F_1=$64.52\%.

\subsubsection{Selection of the Threshold for the Present VAD}
We initially performed a nested-grid search to search for the optimal threshold (denoted as $\theta$) for the present RPCA-based VAD in the KP data set. The result is shown in Figure \ref{fig:thr_SA_response}. The SA of 62.76\% in the main result of Section \ref{subsec:SA_in_KP} was achieved by setting the VAD threshold $\theta=$1.88, which maximized the $F_1$-measure in the KP data set. Then, we reused this fixed threshold for the experiment with EP in Section \ref{subsec:WA_in_EP}.  

\subsubsection{The effects of VAD false positives ({\it fp}) versus false negatives ({\it fn}) in the proposed system}
\label{ssec:fp_fn}
In some classification problems, where the number of true/false examples are extremely unbalanced, giving more weight to {\it precision} or {\it recall} could be obviously advantageous to the end user---in spam mail classification and cancer classification, to name a few cases. In music recordings, indeed, the number of voice inactive-frames is generally larger than that of voice-active frames. In this regard, the existing ASR-based {\it lyrics-to-audio} systems\cite{fujihara2006automatic,fujihara2011lyricsynchronizer,mauch2012integrating} required fixing the VAD threshold manually to adapt to the middle ASR classifier. Additionally, the final threshold was controlled adaptively by {\it rule-of-thumb}\cite{otsu1975threshold} using a histogram of VAD decisions that were made initially. Necessarily, these processes require an {\it a priori} step with respect to the preferred balance of {\it precision} (related to {\it fp}) and {\it recall} (related to {\it fn}) to maximize the final alignment accuracy.

In our case, the best SA (marked with *) in Figure \ref{fig:pr_curve} was achieved using the VAD threshold ($\theta$) for the maximum $F_1$, while the other compared thresholds for the maximum $F_{0.5}$ and $F_{2}$ produced similarly lower SAs (indicated as $\circ$ and $\triangle$). Although our sample size was limited, this result indicated that the $fp$ and $fn$ rate of VAD affected the performance of the present system almost equivalently, and we could see that the VAD output with the best $F_1$-measure could more generally lead us to the best SA performance.

\section{Conclusions}
\label{sec:print}
We presented a new approach for {\it lyrics-to-audio} alignment in popular music recordings. In comparison with the existing ASR-based approaches, the present method differed primarily in that it was based on unsupervised learning, and the fundamental idea was to use the patterns of vowel repetition observed in both audio and text as key features. In this way, the proposed system was shown to work without the need for any pre-constructed models for phoneme recognition, singing/singer adaptation, or language.

The proposed WS-NMF+CTW consisted of two sequentially organized methods: 
WS-NMF, a structured factorization that approximated the discriminative vowel subspace, could produce a sparse matrix representing the patterns of vowel activation from a self-similarity matrix. Then, CTW estimated the optimal spatiotemporal transformation for alignment of the sparse matrix and another matrix from the proposed text processor.

In experiments, we used an existing singing voice separation method and its derivative voice activity detection method as preprocessors. In experiment \ref{subsec:SA_in_KP}, the present system achieved syllable-level accuracy of 62.76\% with the K-pop data set, and outperformed the other unsupervised methods we compared. In experiment \ref{subsec:WA_in_EP}, the present system showed 32.16\% higher word-level accuracy than the compared existing ASR-based system under the same experimental conditions with the English-pop data set. In additional experiments, we investigated three related issues: validation of the previously made assumption that used only vowels, the benefits of combining WS-NMF with CTW, and the effects of VAD error on the performance of the present system.

Although the present study showed promising results, we did not discover a way of fully integrating the voice separation, nor a semi-supervised scenario using known phonetic distances. Thus, these are issues for future work.



\bibliographystyle{IEEEtran}
\bibliography{ref_lyric2audio,ref_snmf_seg}


\begin{IEEEbiography}[]{Sungkyun Chang (M'14)}
received the Master of Science in Engineering degree from Seoul University, Seoul, Korea, in 2013, and the Bachelor of Music degree in music composition from Seoul University, Seoul, Korea, in 2008.\\ 
He is currently a researcher at the Music and Audio Research Group. His previous research interests focus on music information retrieval, with a strong emphasis on audio signal processing, machine learning and computational music theory, such as singing voice onset detection and tonal tension prediction from music signals. His other research interests include deep learning inspired by cross-modal processing in the human brain. 
\end{IEEEbiography}

\begin{IEEEbiography}{Kyogu Lee} received the PhD in computer-based music theory and acoustics from Stanford University.\\
He is currently an associate professor at Seoul National University and leads the Music and Audio Research Group. His research focuses on signal processing and machine learning techniques applied to music and audio. 
\end{IEEEbiography}

\vfill

\end{document}